\documentclass[pra,twocolumn,showpacs,preprintnumbers,amsmath,amssymb,superscriptaddress,aps,10pt]{revtex4-1}

\usepackage{graphicx}
\usepackage{dcolumn}
\usepackage{amsmath}
\usepackage{textcomp}
\usepackage{color}
\usepackage{braket}

\usepackage[caption=false]{subfig}

\begin{document}

\title{Ultrafast charge redistribution in small iodine containing molecules}

\author{M. Hollstein}
\email{Maximilian.Hollstein@uni-hamburg.de}
\affiliation{Department of Physics, University of Hamburg, Jungiusstra{\ss}e 9, D-20355 Hamburg, Germany}
\author{K. Mertens}
\email{Karolin.Mertens@desy.de}
\affiliation{Department of Physics, University of Hamburg, Luruper Chaussee 149, D-22761 Hamburg, Germany}
\author{N. Gerken} 
\affiliation{Department of Physics, University of Hamburg, Luruper Chaussee 149, D-22761 Hamburg, Germany}
\author{S. Klumpp}
\affiliation{Department of Physics, University of Hamburg, Luruper Chaussee 149, D-22761 Hamburg, Germany}
\affiliation{DESY, Notkestra{\ss}e 85, D-22607 Hamburg, Germany}
\author{S. Palutke}
\affiliation{Department of Physics, University of Hamburg, Luruper Chaussee 149, D-22761 Hamburg, Germany}
\author{I. Baev}
\affiliation{Department of Physics, University of Hamburg, Luruper Chaussee 149, D-22761 Hamburg, Germany}

\author{G. Brenner}
\affiliation{DESY, Notkestra{\ss}e 85, D-22607 Hamburg, Germany}
\author{S. Dziarzhytski}
\affiliation{DESY, Notkestra{\ss}e 85, D-22607 Hamburg, Germany}

\author{W. Wurth}
\affiliation{Department of Physics, University of Hamburg, Luruper Chaussee 149, D-22761 Hamburg, Germany}
\affiliation{DESY, Notkestra{\ss}e 85, D-22607 Hamburg, Germany}

\author{D. Pfannkuche}
\affiliation{Department of Physics, University of Hamburg, Jungiusstra{\ss}e 9, D-20355 Hamburg, Germany}

\author{M. Martins}
\affiliation{Department of Physics, University of Hamburg, Luruper Chaussee 149, D-22761 Hamburg, Germany}


\date{\today}


\begin{abstract}
The competition between intra molecular charge redistribution and fragmentation has been studied in small molecules containing iodine by using intense ultrashort pulses in the extreme ultraviolet regime (XUV). We show that after an element specific inner-shell photoionization of diiodomethane (CH$_2$I$_2$) and iodomethane (CH$_3$I), the induced positive charge is redistributed with a significantly different efficiency. Therefore, we analyze ion time-of-flight data obtained from XUV-pump XUV-probe experiments at the Free Electron Laser in Hamburg (FLASH). Theoretical considerations on the basis of \textit{ab initio} electronic structure calculations including correlations relate this effect to a strongly molecule specific, purely electronic charge redistribution process that takes place directly after photoionization causing a distribution of the induced positive charge predominantly on the atoms which exhibit the lowest atomic ionization potential, i.e, in the molecules considered, the iodine atom(s).
As a result of the very different initial charge distributions, the fragmentation timescales of the two molecules experimentally observed are strikingly different.  \end{abstract}

\maketitle

Charge rearrangement between atomic centers is vital for the formation and breaking of chemical bonds in molecules. In particular, ionization induced by photoabsorption or electron impact leads very often to fragmentation of molecules.  In this context, the understanding of how charge is redistributed directly after the ionization is central, since the arising charge distribution triggers the subsequent nuclear motion.  
For inner-shell ionization, photo excitation is element-specific and results in localized electron holes which can decay via electron emission.  This decay results in (delocalized) valence vacancies and allows for ultrafast inter-atomic charge redistribution which is only limited by  atomic separation caused by nuclear motion. In experiments, this process has been observed to follow inner-shell photoionization of methylselenol \cite{PhysRevLett.110.053003} and ethylselenol \cite{0953-4075-46-16-164031} and it has been found that its spatial range is well estimated by the classical over-the-barrier model \cite{PhysRevLett.113.073001,Erk18072014}. Recently, the  occurrence of this process has been demonstrated for distances between molecular fragments up to 20 angstroms \cite{Erk18072014}. 

In this Letter, we present a combined theoretical and experimental effort showing that this electronic charge redistribution and the subsequent fragmentation are highly dependent on the specific molecule. In particular, the first electronic rearrangement and the initial stages of the nuclear motion extremely depend on the molecule's size and structure. Experimentally, we investigate the charge redistribution following inner-shell photoionization of CH$_3$I and CH$_2$I$_2$ using an XUV-pump-XUV-probe scheme. Here, the pump pulse creates an inner-shell iodine-4d vacancy and the probe pulse probes the status of the fragmentation that is triggered by the initial charge distribution. Theoretically, we investigate this initial charge distribution resulting from the decay of the iodine-4d vacancy triggering the fragmentation dynamics probed in the experiment. With this joint experimental and theoretical approach, we can show that in these two similar molecules, which differ only by a single atom, the positive charge induced by XUV photoionization is redistributed with strikingly different efficiency. As a result, also the subsequent fragmentation of the two molecules is very different. Notably, the overall remarkable good agreement between experimental results and theoretical considerations based on {\it ab initio} electronic structure calculations  indicates that we can reveal the relevant processes that are not only relevant in the molecules considered but can be expected to crucially determine the evolution of molecules following XUV ioniziation in general.

\begin{figure}
\includegraphics[width=0.5\textwidth]{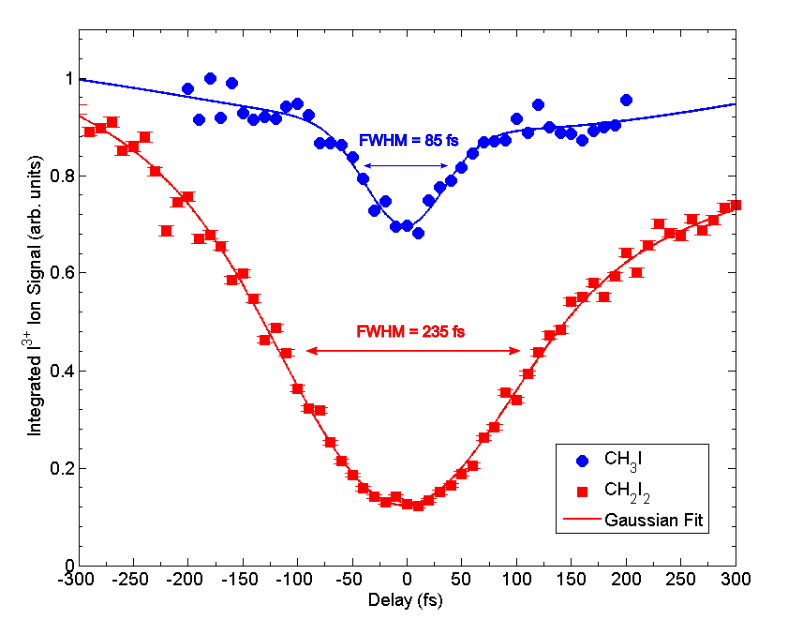}
  \caption{\label{fig:iodinetimedelay} The integrated I$^{3+}$ signal in arbitrary units arising after the interaction of CH$_3$I and CH$_2$I$_2$ molecules with the pump and the probe pulse in dependence of the pump-probe time delay. The reduction of the I$^{3+}$ signal for vanishing time delays with respect to large time delays indicates charge redistribution following XUV photoionization that occurs as long as the molecular fragments are in close vicinity. For CH$_2$I$_2$, charge redistribution is appearently much more efficient as in CH$_3$I as indicated by the fact that the I$^{3+}$ signal almost vanishes for vanishing time delays whereas for CH$_3$I in the same situation  a significant signal ($\sim 0.7$) is still observable.}
\end{figure}

 The experiment has been realized at the PG2 beamline \cite{martins2006b, wellhoefer2007a,doi:10.1080/09500340.2011.588344} at the free-electron laser at DESY in Hamburg (FLASH) \cite{ackermann2007a,1367-2630-11-2-023029}. The time resolved XUV-pump XUV-probe studies have been performed at 82.7 eV photon energy with average pulse energies of 40 $\mu$J and a focal spot size of about 50 $\mu$m using the split-and-delay unit (SDU) of the PG2 beamline, which is based on the grazing incidence Mach-Zehner geometry \cite{flashsdu2010}. It can split a XUV pulse into two and induces a jitter-free adjustable delay up to 5.1 ps between the splitted parts of the beam. The FEL pulse duration was in the range of 80 fs $\pm$ 30 fs for the CH$_3$I and 100 fs $\pm$ 30 fs for the CH$_2$I$_2$, which has been estimated by measuring the electron bunch length by the LOLA setup \cite{PhysRevSTAB.17.120702}.
FLASH has been run in the multi-bunch mode generating 40 electron bunches with a bunch separation of 10 $\mu$s and a bunch train repetition rate of 10 Hz. This mode allows us to collect a sufficient number of single shot spectra at each pump-probe time delay. The pulse energies, measured by the FLASH gas monitor detectors (GMD) \cite{tiedtke:094511}, and the single shot spectra have been recorded on a shot-to-shot basis using a fast transient recorder (Acqiris ADC DC282) and the FLASH fast data acquisition system (DAQ). This guarantees a perfect synchronization of all measured properties and it allows a post-measurement sorting of the single-shot spectra according to the photon exposure depending on the fluctuating single-shot intensity produced by the Self-Amplified Spontaneous Emission (SASE) process \cite{ackermann2007a}. The created charged fragments of the gas phase molecules have been detected by ion time-of-flight (iTOF) spectroscopy. 
A single count discrimination scheme is applied to the iTOF signal. The mass-to-charge spectrum is obtained by summing over all single shot spectra and by sorting according to GMD values. The photon density was carefully chosen such as that only one and two photon processes are possible. The highest charge state found, I$^{3+}$, is the first charge state which can be reached with two subsequent photoionization processes.  Fig. \ref{fig:iodinetimedelay} shows the integrated  I$^{3+}$ charge state signal in dependence of the time delay of the two split beams. A pronounced time-dependence can be observed.
 At $t = 0$ fs, the splitted beams arrive simultaneously at the molecule resulting in a decreased I$^{3+}$ signal for both molecules as compared to the signal for long delay time. With increasing delay between the two FEL pulses the signal increases. While for CH$_2$I$_2$ the I$^{3+}$ signal goes almost to zero for zero delay, for CH$_3$I a reduction by only $\sim$ 20 \% - 30 \% is observed. A Gaussian fit to the data yields a characteristic timescale for the increase of the  I$^{3+}$  signal. For CH$_3$I, this can be estimated to 85 fs $\pm$ 20 fs while CH$_2$I$_2$ shows a much longer timescale of 235 fs $\pm$ 20 fs. This behavior is a clear indication for ultrafast charge redistribution which can be sensitively witnessed in the I$^{3+}$ channel in iodine-containing molecules \cite{PhysRevLett.113.073001}. Using photon energies in the vicinity of the iodine 4d giant resonance, both pump and probe pulse create predominantly iodine-4d vacancies which decay within a few femtoseconds via electron emission \cite{doi:10.1063/1.463468,krikunova:024313}.  As long as the molecule is intact, this Auger decay, which results in dicationic states, is typically accompanied with the redistribution of positive charges throughout the whole molecule. This initiates the dissociation. The second Auger decay initiated by the probe pulse repeats this process. However, if the fragments reach a critical distance, charge redistribution is no longer possible and all additional positive charges created by the second Auger decay remain on the separated iodine atoms. This leads to an increase of the iodine I$^{3+}$ signal with increasing pump probe delay. 

\begin{figure}

\subfloat[\label{fig:partial_charges_and_two_hole_population_analysis_CH2I2} CH$_2$I$_2$]{%
      \includegraphics[width=.5\linewidth]{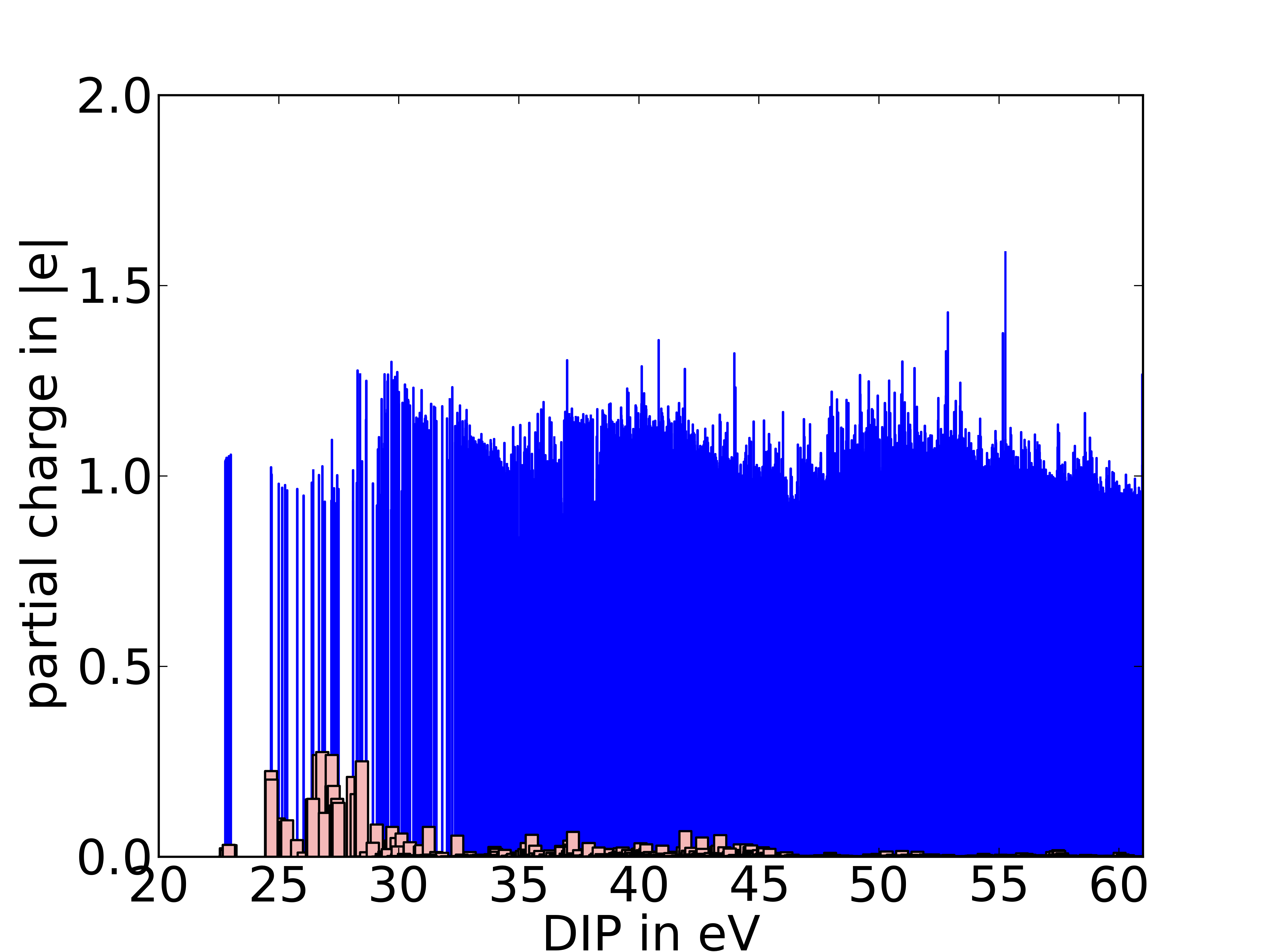}
    }
    \subfloat[\label{fig:partial_charges_and_two_hole_population_analysis_CH3I} CH$_3$I]{%
      \includegraphics[width=.5\linewidth]{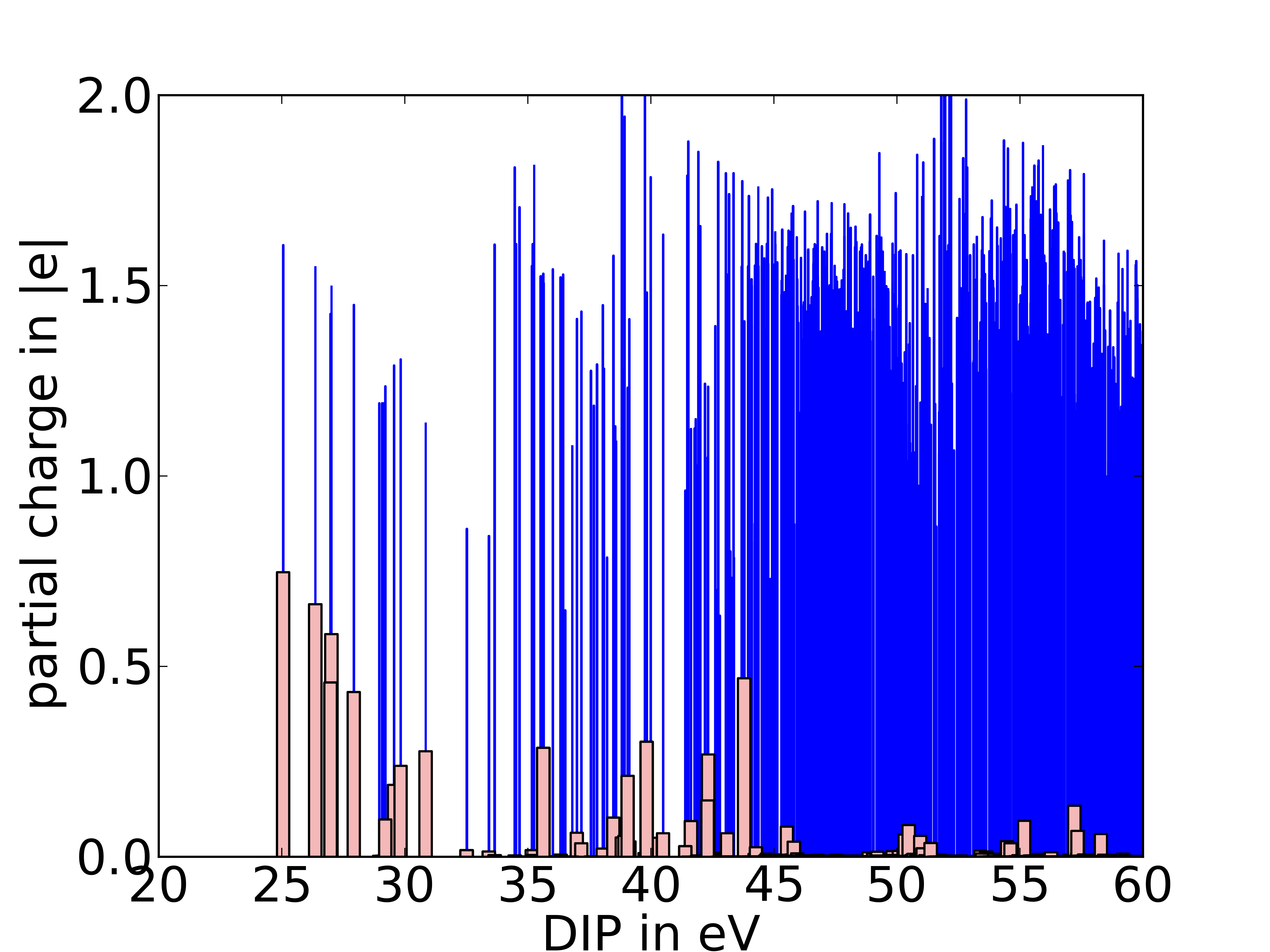}
    }
    \caption{
\label{fig:partial_charges_and_two_hole_population_analysis_CH2I2_CH3I} The iodine partial charges (thin, blue sticks) for the dicationic eigenstates and the I$^{+2}$ two-hole populations (thick, rose sticks). For the dicationic eigenstates below 60 eV, it is clearly visible that the iodine partial charges of the dicationic eigenstates of CH$_3$I are shifted, compared to CH$_2$I$_2$, to higher charges.}

\end{figure}

To understand the different efficiency of charge rearrangement in CH$_3$I compared  to CH$_2$I$_2$, we carried out numerical calculations concerning the first electronic charge redistribution.  For two reasons, this first process is central: first, it is most efficient since it occurs when the atoms constituting the molecules are still in their initial positions in close vicinity and second, it determines all subsequent processes by setting the stage for the ensuing fragmentation dynamics. Therefore, we determine final dicationic states that are populated after the first secondary electron emission using highly correlated \textit{ab initio} electronic structure calculations. The dicationic eigenstates were determined by a multi-reference configuration interaction approach relying on the expansion of the dicationic states in terms of the two-hole (2h)  and three-hole-one-particle (3h1p) configurations constructed  with respect to the Hartree-Fock ground state. The transition probabilities $p_{\Psi}$ to these dicationic eigenstates $\Psi$ were obtained by following the procedure described in Ref. \cite{Tarantelli1991} known as two-hole population analysis. For this, owing to the dominance of intra-atomic transition matrix elements, the transition probabilities are approximated to be proportional to the I$^{+2}$ two-hole populations: \begin{equation}p_{\Psi} \sim \sum_{i<j}\mid\langle i,j|\Psi\rangle\mid^2\end{equation} Here, $|i,j\rangle$ denotes a two-hole state $|i,j\rangle=c_{\phi_i} c_{\phi_j} |\Phi_0\rangle$ where $c_{\phi_i}$ and $c_{\phi_j}$ are annihilation operators that annihilate electrons from atomic iodine spin orbitals $\phi_i$ and $\phi_j$ obtained from L\"owdin's symmetric orthogonalization. $|\Phi_0\rangle$ denotes the Hartree-Fock ground state. The Hartree-Fock orbitals and the one-particle integrals of the kinetic energy and the core potentials needed for the construction of the Hamiltonian matrix, are taken from the quantum chemistry software MOLCAS \cite{Karlstrom2003222}. For the iodine atoms and the carbon atom, we employed effective core potentials and associated valence basis sets from Ref. \cite{pseudopot} whereas for the hydrogen atoms 6-31G basis sets were used. The Coulomb matrix elements were calculated by the LIBINT library \cite{libint}. The molecular geometries were obtained by MOLCAS on the Hartree-Fock SCF wavefunction level. \newline In order to get insight into the charge distributions associated with the populated dicationic eigenstates, we determined the partial charges of the iodine atoms in the molecules by L\"owdin population analysis. Both, partial charges and I$^{+2}$ two-hole populations are shown in Fig. \ref{fig:partial_charges_and_two_hole_population_analysis_CH2I2_CH3I} for the part of the dicationic spectrum that can be populated after the electron emission (i.e. dicationic states with a double ionization potential less than the iodine 4d ionization potential i.e. $\lesssim 60$ eV). For the relevant part of the dicationic spectrum it is clearly visible that the iodine partial charges of the dicationic eigenstates of CH$_3$I are shifted, compared to CH$_2$I$_2$, to higher charges. Moreover, weighted by the I$^{+2}$ two-hole populations, average iodine partial charges result in $0.98|e|$ for each of the two iodine atoms in CH$_2$I$_2$ and  in $1.39|e|$ for the single iodine atom in CH$_3$I. This indicates that charge can be more efficiently redistributed in CH$_2$I$_2$ than in CH$_3$I already after the first photoionization. That is, assuming a negligible transfer of two electrons associated with the molecular Auger decay, one can estimate the transfer probability of a single electron in CH$_3$I to $\sim 0.6$ whereas in CH$_2$I$_2$ this transfer probability can be estimated to be nearly 1.  

\begin{figure}
\includegraphics[width=0.5\textwidth]{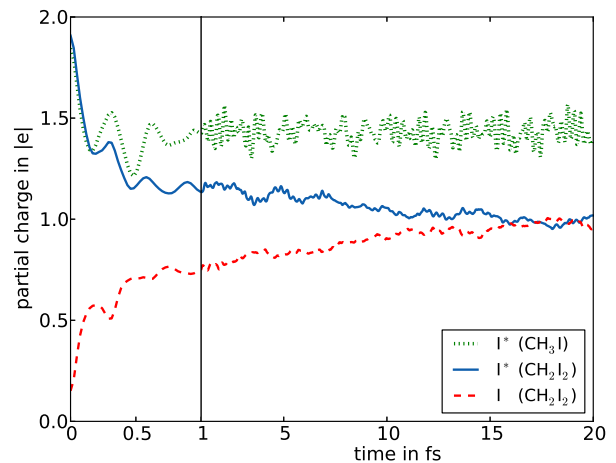}
\caption{\label{fig:dynamics_subsequent_to_auger_decay} The time-dependent partial charges of the iodine atoms after the sudden removal of two atomic iodine valence electrons averaged over all possible initial states (i.e. all possible initial iodine double valence hole states). Here, I$^*$ denotes the iodine site where the two holes were initially situated. The atomic charges are obtained from a L\"owdin population analysis. In both molecules, most charge appears to be redistributed within less than 1 femtosecond. }
\end{figure}

In the following, we point out that these considerably different charge distributions can arise even before a significant nuclear motion can take place. To estimate the timescale for the interatomic charge redistribution, we consider the limiting situation where two valence electrons are suddenly removed from one of the iodine atoms in CH$_2$I$_2$ and respectively the iodine atom in CH$_3$I. Concretely, we considered the time-evolution of the iodine two-hole states:  $|\psi \rangle$  
\begin{equation}\label{initial_state}|\psi \rangle = c_{\phi_i} c_{\phi_j} |\Phi_0\rangle\end{equation} where $c_{\phi_i}$ and $c_{\phi_j}$ are annihilation operators that annihilate electrons from atomic iodine spin valence orbitals $\phi_i$ and $\phi_j$ obtained from L\"owdin's symmetric orthogonalization. As for the dicationic eigenstates, we expand these time-evolved states in terms of two-hole and the three-hole-one-particle configurations. The time evolution of the iodine partial charges averaged over all possible initial iodine two-hole states is shown in Fig. \ref{fig:dynamics_subsequent_to_auger_decay}, for both CH$_2$I$_2$ and CH$_3$I.
 There, one can observe that most of the positive charge can be redistributed within only a few hundred attoseconds---i.e. clearly before a significant nuclear motion can take place that could prevent the electron charge redistribution.  
 A complete charge equalization of the iodine atoms in CH$_2$I$_2$ is reached after $\sim $ 15 fs, i.e., still within a time span during which in particular the heavy iodine nuclei move only little. In both molecules, charge state distributions arise that are very similar to those obtained from the two-hole population analysis. Hence, already after the first photoionization, before a nuclear displacement can occur that can prevent the electronic charge redistribution, charge is more efficiently redistributed in CH$_2$I$_2$ than in CH$_3$I.
Notably, the induced positive charge is redistributed in both molecules mostly on the iodine atom(s)---i.e. in CH$_2$I$_2$ on the two iodine atoms  and in CH$_3$I, respectively, the single iodine atom.  Remarkably, for CH$_2$I$_2$, this effect is not only due to the preferable population of dicationic states with large I$^{+2}$ two-hole populations but also due to the population of 3h1p configurations. That is, the average partial charges obtained from the two-hole population analysis are significantly smaller when the 3h1p configurations are excluded---i.e. for CH$_2$I$_2$, without the 3h1p configurations, the iodine partial charges would be $0.88|e|$  instead of $0.98|e|$.
This indicates that the excitation of an electron to initially unoccupied orbitals allows a 'relaxation' of the positive charge on the iodine atoms which have the lowest atomic ionization potential in the molecules considered and fosters spatial separation of the induced positive charges. Notably, the charge equalization in CH$_2$I$_2$ is completed only after $\sim $ 15 fs. This finding shows that the inter-atomic electronic charge redistribution may exceed the electron emission taking place within a few femtoseconds \cite{:/content/aip/journal/jcp/97/11/10.1063/1.463468}. This indicates that in CH$_2$I$_2$, the Auger decay may prepare the resulting dication in a (partially) coherent state. We note that this initial electronic charge redistribution that exceeds the electron emission associated with the molecular Auger decay might be observed using attosecond transient absorption spectroscopy. However, from the time-dependent calculations presented above one can also conclude that even if in the course of the molecular Auger decay dicationic eigenstates are populated (partially) coherently, very rapidly a quasi stationary charge distribution arises---i.e. large charge oscillations appear to be suppressed. This effect appears to be due to the dephasing as a result of the population of many dicationic eigenstates with very different energies and comes in addition to the dephasing induced by nuclear motion. Therefore, a stationary charge distribution can be expected to arise already $\sim$ 15 fs after the electron emission.\newline


With regard to the experiment, we can draw the following conclusions from the above calculations: the first electronic charge redistribution is essentially completed before a significant nuclear displacement occurs. While in CH$_2$I$_2$, the whole positive charge is equally distributed on the two iodine atoms, the charge redistribution efficiency in CH$_3$I is only 0.6. Assuming a negligible efficiency for I$^{3+}$ resulting from single photoionization followed by a cascade of Auger decays, the almost complete charge redistribution in CH$_2$I$_2$ prevents the occurrence of the I$^{3+}$ ions also in two subsequent photoionization processes. The finite I$^{3+}$ signal for CH$_3$I at zero delay witnesses the probability for single charge transfer in the two subsequent ionization processes. To estimate the charge distributions that can occur after a second photo ionization, we also performed a calculation similar to those presented above for the initially neutral molecules, however, this time for the dication. Provided the molecule is still intact when the second photon is absorbed, we can estimate the iodine partial charge of CH$_3$I, which arises after the second photo ionization and subsequent Auger decay, to 2.66 $|e|$. When neglecting extreme iodine charge states such as I$^{1+}$ or lower and I$^{4+}$ and larger, this result allows one to estimate the probability for the transfer of two electrons from the iodine atom to the CH$_3$ group---provided the second photon is absorbed when the molecule is still intact---to 0.34. This effect provides one source for a finite I$^{3+}$ signal at zero delay (see Fig. \ref{fig:iodinetimedelay}).
 
We now turn to the characteristic timescale for the increase of the I$^{3+}$ signal observed in the experiment. It indicates the number of processes with incomplete charge transfer following the ionization by the probe pulse which is caused by the repulsive motion of the respective ions. For CH$_2$I$_2$, the positive charges are distributed on the comparatively heavy and more separated iodine atoms. This results in slowed fragmentation dynamics.  According to the over-the-barrier model \cite{PhysRevA.21.745,0022-3700-19-18-021,PhysRevLett.113.073001}, charge transfer is possible for distances up to 9.8 au which the two ions reach after $\sim$ 145 fs. Hence, for overlapping XUV pulses ($t = 0$), the two singly charged iodine ions are close enough to allow charge redistribution for the whole exposure of the two pulses. The strongly suppressed I$^{3+}$ signal for vanishing time delays is a clear evidence for a second redistribution after a second photoionization. At large time delays ($t \gg 145 $ fs), a charge redistribution after a second photoionization is suppressed since the fragments  are too far apart when a second electron is photoionized. 
The I$^{3+}$ signal at large time delays evidences the secondary photoionization of an isolated iodine ion.\newline
Compared to CH$_2$I$_2$, the over-the-barrier model predicts for CH$_3$I vanishing charge transfer between the iodine and the carbon atom at a distance of 7.7 au. The lighter carbon atom reaches this distance already after $\sim$ 40 fs---i.e. within a quarter of the characteristic timespan of the iodine ions in CH$_2$I$_2$. The good agreement of this estimate with the experimental results underlines our interpretation of the relevant processes.
Due to the fast fragmentation dynamics of CH$_3$I, for CH$_3$I, additionally to the comparably low charge transfer probability ($\sim 0.6$), the suppression of charge redistribution due to fragmentation is also a source for a finite I$^{3+}$ signal at zero time delay. Considering these two effects which can cause a finite I$^{3+}$ signal, the reduction of the I$^{3+}$ signal for vanishing time delays with respect to large time delays can be estimated on basis of a simple model (see supplemental material) to 0.16 $\pm_{0.12}^{0.06}$. In excellent agreement with the experimental findings, this estimate reflects the circumstance that in CH$_3$I, charge is much less efficient redistributed. Notably, within the error tolerances, this estimate is in accordance with the experimental value which is found to be between 0.2 to 0.3.\newline\newline
To conclude, we have demonstrated that in two similar molecules CH$_2$I$_2$ and CH$_3$I, which differ only by a single atom, the positive charge induced by XUV photoionization is redistributed with substantially different efficiency. With the help of theoretical considerations based on {\it ab initio} electronic structure calculations which are in good agreement with the experimental results, we pointed out that this can be related to the initial, purely electronic charge redistribution that takes place within a few femtoseconds after photoionization. It has been demonstrated that thereby locally induced positive charge appears to be predominantly distributed on the atoms which exhibit the lowest atomic ionization potential in the molecules, i.e, here the iodine atom(s). 
This process leads to very different initial charge distributions triggering the respective fragmentation dynamics which in turn limits charge redistribution in subsequent photoionization processes.

\section*{Acknowledgment}

The work was funded by the DFG and performed within the framework of SFB925, projects A3, A5 and B2 and MA2561/4-1. We thank Michael Meyer for his continuous support. The help of the DESY staff operating FLASH in an excellent way is gratefully acknowledged.
MH and DP are grateful to Robin Santra for helpful discussions.
\bibliography{lit}

\end{document}


\title{Ultrafast charge redistribution in small iodine containing molecules - Supplemental material}

\author{M. Hollstein}
\email{Maximilian.Hollstein@uni-hamburg.de}
\affiliation{Department of Physics, University of Hamburg, Jungiusstra{\ss}e 9, D-20355 Hamburg, Germany}
\author{K. Mertens}
\affiliation{Department of Physics, University of Hamburg, Luruper Chaussee 149, D-22761 Hamburg, Germany}
\author{N. Gerken} 
\affiliation{Department of Physics, University of Hamburg, Luruper Chaussee 149, D-22761 Hamburg, Germany}
\author{S. Klumpp}
\affiliation{Department of Physics, University of Hamburg, Luruper Chaussee 149, D-22761 Hamburg, Germany}
\affiliation{DESY, Notkestra{\ss}e 85, D-22607 Hamburg, Germany}
\author{S. Palutke}
\affiliation{Department of Physics, University of Hamburg, Luruper Chaussee 149, D-22761 Hamburg, Germany}
\author{I. Baev}
\affiliation{Department of Physics, University of Hamburg, Luruper Chaussee 149, D-22761 Hamburg, Germany}

\author{G. Brenner}
\affiliation{DESY, Notkestra{\ss}e 85, D-22607 Hamburg, Germany}
\author{S. Dziarzhytski}
\affiliation{DESY, Notkestra{\ss}e 85, D-22607 Hamburg, Germany}

\author{W. Wurth}
\affiliation{Department of Physics, University of Hamburg, Luruper Chaussee 149, D-22761 Hamburg, Germany}
\affiliation{DESY, Notkestra{\ss}e 85, D-22607 Hamburg, Germany}

\author{D. Pfannkuche}
\affiliation{Department of Physics, University of Hamburg, Jungiusstra{\ss}e 9, D-20355 Hamburg, Germany}

\author{M. Martins}
\affiliation{Department of Physics, University of Hamburg, Luruper Chaussee 149, D-22761 Hamburg, Germany}

\date{\today}

\pacs{}

\begin{abstract}

\end{abstract}
\maketitle
Here, we provide additional information concerning the theoretical estimate of the finite I$^{3+}$ signal for CH$_3$I at vanishing time delay between pump and probe pulse. 
Therefore, we present a simple model of the relevant processes that result in I$^{3+}$ ions.
First, we note that according to the two-hole population analysis, following the first electron emission, predominantly dicationic eigenstates below the triple ionization threshold at $\sim 52 $ eV (determined on the level of Koopmans theorem applied to the dication) are populated with a probability of $\sim$ 0.9. This indicates that processes involving the emission of two electrons following the iodine-4d photoionization are of minor importance. Therefore, in our model, we neglect these processes - i.e. we do not include Auger cascades into our considerations. 

In the experiment, the pulse durations are on the order of 80 fs. This means that for zero time delay, both photons are absorbed within $\sim $80 fs. Notably, this time interval is by no means short with respect to the timescale of the nuclear motion of the ionic fragements of CH$_3$I.  For instance, driven by their Coulomb repulsion, a singly charged iodine ion and a singly charged carbon ion can enlarge their distance within only 40 fs to 7.7 au so that, according to the over-the-barrier model, from then on, no further charge transfer between them can occur. Hence, a suppression of a charge transfer following the absorption of a second photon and the corresponding creation of I$^{3+}$ ions can also occur for vanishing time delays. For a direct comparison of the experimental data with the theoretical predictions presented in the paper, this effect has to be taken into account. In order to estimate it's relevance, we determine, given that two photons are absorbed, the conditional probability that the two photons are absorbed within a time interval so that charge redistribution between the iodine and the carbon site can follow both ionization events. For simplicity, we assume that the probability for photoionization events occuring within the pulse during time intervals $[t,t +dt]$ to be time independent. Considering the large uncertainties associated with the width of the pulses ($\pm 30 $ fs), an more accurate consideration of the pulse (form) appears unreasonable. With this, given that two photons are absorbed, the conditional probability for an event in which the second photon is absorbed in time, so that charge redistribution is still possible, can be estimated by:\begin{equation}p^{t = 0}_{\text{no-frag}} (T_{frag},T) = (2 -\frac{T_{frag}}{T}) \frac{T_{frag}}{T}\end{equation} where $T_{frag}$ denotes the critical temporal distance between the two photoionization events from which on charge redistribution cannot occur any more. T denotes the pulse length. In the experiment, $\frac{T_{frag}}{T}$ is on the order of $0.5 \pm 0.38$ so that here, $p^{t = 0}_{\text{no-frag}}$ is $\sim 0.75 \pm_{0.16}^{0.21}$. Hence, the probability that the second photon is absorbed when a charge transfer is not possible any more due to the separation of the fragments, i.e., $p^{t = 0}_{\text{frag}} = 1 - p^{t = 0}_{\text{no-frag}} = 0.25 \pm_{0.21}^{0.16}$ is not negligible. \newline For large time delays $(t \gg 40 \text{fs})$, when the pulses do not overlap significantly, one finds: \begin{equation}p^{t \gg 40 fs}_{\text{no-frag}} (T_{frag},T) = \frac{1}{2}(2 -\frac{T_{frag}}{T}) \frac{T_{frag}}{T}\end{equation}

\begin{figure*}
\includegraphics[width=1.0\textwidth]{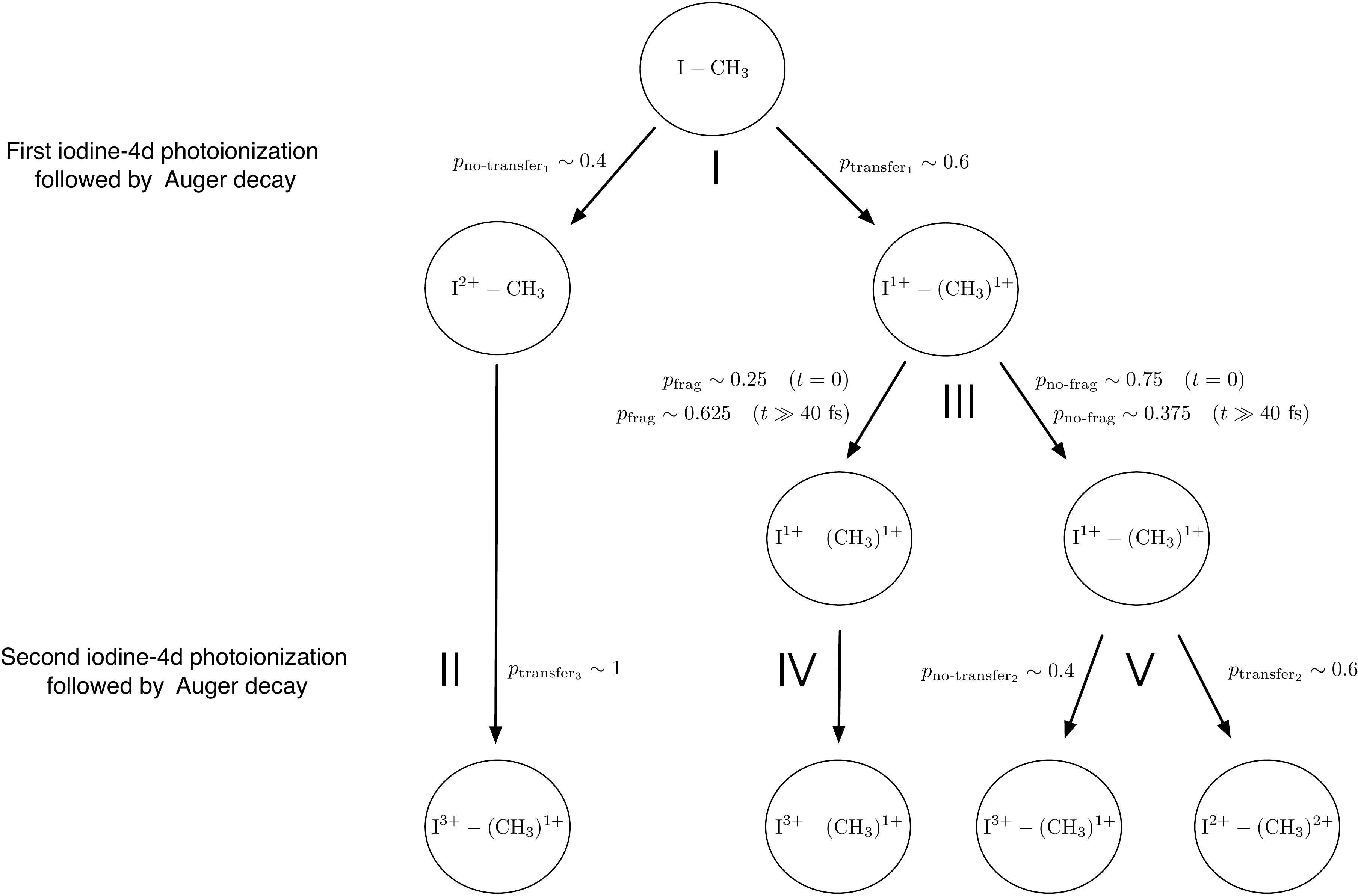}
  \caption{\label{fig:probability_tree} Note, a hyphen between I$^{\text{X}+}$ and the (CH$_3$)$^{\text{Y}+}$  indicates that charge transfer between those ionic fragments can still occur. \newline\newline \textbf{I} In the course of the molecular Auger decay following the first iodine-4d photoionization, positive charge is transferred from the iodine atom to the CH$_3$-group with a probability of 0.6 (see the paper).  \newline\newline \textbf{II} In the experiment, no I$^{4+}$ ions were observed. Therefore, we conclude that the transfer probability $p_{\text{transfer}_3}$  is $\sim 1$.    \newline\newline \textbf{III} Following the first photoionization and molecular Auger decay, the fragmentation of the molecule starts.  With the probability of $p_{frag}$, the distance between the fragments enlarges so that no charge redistribution subsequent to a second photoionization is possible.\newline\newline \textbf{IV} The second photon is absorbed when the iodine ion is separated from the other fragments and positive charge induced by a second photoionization and ensuing Auger decay remains on the isolated iodine atom.\newline\newline \textbf{V} The second photon is absorbed when the iodine is still in the vicinity of the other fragments and charge redistribution is still possible. From our calculation, we could estimate the probability of the first charge transfer to 0.6 and the probability of two ensuing transfers---in the situation where the molecule is still intact---to 0.34. This indicates that also for the second photo absorption, the charge transfer probability $p_{\text{transfer}_2}$ is on the order of 0.6.}
\end{figure*}

With this, a probability tree can be created (see Fig. \ref{fig:probability_tree}) concerning the relevant processes and the reduction of the I$^{3+}$ signal for vanishing time delays with respect to large time delays ($t \gg 40 $ fs) can be estimated to 0.16 $\pm_{0.12}^{0.06}$  ---in excellent agreement with the experimental findings, this estimate reflects the circumstance that in CH$_3$I, charge is much less efficient transferred from the iodine atom to it's molecular environment than in CH$_2$I$_2$. Notably, within the error tolerances, this estimate is in accordance with the experimental value which is found to be between 0.2 to 0.3.
